\documentclass[12pt,twoside]{article}
\usepackage{wrapfig}
\usepackage{graphicx}
\usepackage{cmp2e}
\title[Non-universal critical behaviour of a mixed-spin Ising model]%
{Non-universal critical behaviour \\ of a mixed-spin Ising model on \\ the extended Kagom\'e lattice%
\thanks{This work was financially supported under the grants VEGA 1/2009/05 and APVT 20-005204.}}
\author{J. Stre\v{c}ka, L. \v{C}anov\'a}
\address{Department of Theoretical Physics and Astrophysics, \\
Faculty of Science, P. J. \v{S}af\'arik University,  \\
Park Angelinum 9, 040 01 Ko\v{s}ice, Slovak Republic}
\begin{document}

\maketitle

\begin{abstract}

The mixed spin-1/2 and spin-3/2 Ising model on the extended Kagom\'e lattice is solved 
by establishing a mapping correspondence with the eight-vertex model. When the parameter 
of uniaxial single-ion anisotropy tends to infinity, the model system becomes exactly soluble
as the staggered eight-vertex model satisfying the free-fermion condition. The critical points 
within this manifold can be characterized by critical exponents from the standard Ising universality 
class. The critical points within another subspace of interaction parameters, which corresponds 
to a coexistence surface between two ordered phases, can be approximated by corresponding results of 
the uniform eight-vertex model satisfying the zero-field condition. This coexistence surface is bounded 
by a line of bicritical points that have non-universal continuously varying critical indices.

\pacs 75.10.Hk, 05.50.+q, 75.40
\end{abstract}

\section{Introduction}

Investigation of phase transitions and critical phenomena belongs to the most intensively studied 
topics in the equilibrium statistical physics. A considerable progress in the understanding of 
order-disorder phenomena has been achieved by solving planar Ising models that represent valuable
exceptions of exactly soluble lattice-statistical models with a non-trivial critical behaviour \cite{Bax82}. 
Although phase transitions of planar Ising models have already been understood in many respects, 
there are still a lot of obscurities connected with a criticality of more complicated spin systems 
exhibiting reentrant transitions, non-universal critical behaviour, tricritical phenomenon, 
etc. It is worthy to mention, however, that several complicated Ising models can exactly be 
treated by transforming them to the solvable \textit{vertex models}. A spin-1/2 Ising model on 
the union jack (centered square) lattice, which represents a first exactly soluble system exhibiting 
reentrant transitions \cite{Vak66}, can be for instance reformulated as a free-fermion eight-vertex model 
\cite{Wu87}. It should be also pointed out that an equivalence with the vertex models have already provided 
a precise confirmation of the reentrant phenomenon in the anisotropic spin-1/2 Ising models on extended Kagom\'e lattice \cite{Aza87} and centered honeycomb lattice \cite{Die91} as well.

Despite the significant amount of effort, there are only few exactly soluble Ising models consisting 
of mixed spins of different magnitudes, which are usually called also as {\it mixed-spin Ising models}. 
A strong scientific interest focused on the mixed-spin systems arises partly on account of 
much richer critical behaviour they display compared with their single-spin counterparts and 
partly due to the fact that they represent the most simple models of ferrimagnets having 
a wide potential applicability in practice. Using the extended versions of decoration-iteration 
and star-triangle mapping transformations, Fisher \cite{Fis59} has derived exact solutions 
of the mixed spin-1/2 and spin-$S$ ($S \geq 1$) Ising models on the honeycomb, diced and 
decorated honeycomb lattices.  Notice that these mapping transformations were later on further 
generalized in order to account also for the single-ion anisotropy effect. The influence of uniaxial 
and biaxial single-ion anisotropies have precisely been investigated on the mixed-spin honeycomb 
lattice \cite{Gon85} as well as on some decorated planar lattices \cite{Jas98}. With exception 
of several mixed-spin models formulated on the Bethe (Cayley tree) lattices, which can be accurately 
treated within a discrete non-linear map \cite{Sil91} or an approach based on exact recursion equations 
\cite{Alb03}, these are the only mixed-spin planar Ising models with generally known exact solutions, yet. 

One of the most outstanding findings emerging in the phase transition theory is being a non-universal critical behaviour of some planar Ising models, which is in obvious contradiction with the idea of 
universality hypothesis \cite{Gri70}. The mixed spin-1/2 and spin-$S$ Ising model on the union jack lattice \cite{Lip95} represents very interesting system from this viewpoint as it exhibits a remarkable line of bicritical points that have continuously varying critical indices obeying the \textit{weak universality} hypothesis \cite{Suz74}. In the present article, we shall investigate a topologically similar mixed spin-1/2 and spin-3/2 Ising model on the extended Kagom\'e lattice by establishing a mapping correspondence with 
the staggered and uniform eight-vertex models, respectively. In a certain subspace of interaction parameters, the model under investigation becomes exactly soluble as the staggered eight-vertex model satisfying the free-fermion condition \cite{Hsu75}. Even if a non-validity of the free-fermion condition in the rest of parameter space is simply ignored, one still obtains rather reliable estimate of the criticality within  
free-fermion approximation \cite{Fan70}. Finally, the critical points within another subspace 
of interaction parameters can be approximated from the relevant solution of the 
uniform eight-vertex model satisfying the zero-field condition.

The outline of this paper is as follows. In Section 2, a detailed formulation of the model is presented 
and subsequently, the mapping correspondence that ensures an equivalence with the eight-vertex models 
will be derived. The most interesting numerical results for a critical behaviour will be presented and particularly discussed in Section 3. Finally, some concluding remarks are drawn in Section 4.

\section{Formulation}

\begin{figure}
\vspace{-1.2cm}
\centerline{\includegraphics[width=0.5\textwidth]{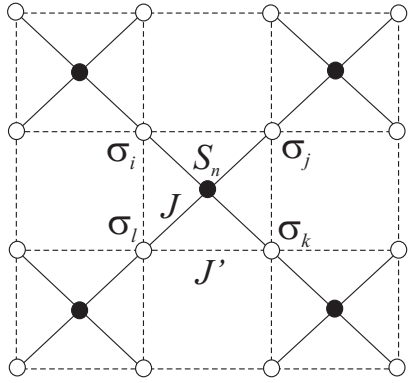}}
\vspace{-1.1cm}
\caption{Diagrammatic representation of the extended Kagom\'e lattice composed of the mixed 
spin-1/2 (empty circles) and spin-3/2 (filled circles) sites, respectively. The solid (broken) 
lines depict the nearest-neighbour (next-nearest-neighbour) interactions.}
\label{fig1}
\end{figure}

Let us begin by considering the mixed spin-1/2 and spin-3/2 Ising model on the extended Kagom\'e 
lattice ${\cal L}$ schematically illustrated in figure \ref{fig1}. The mixed-spin Kagom\'e lattice 
consists of the spin-1/2 (empty) and spin-3/2 (filled circles) atoms placed on the six- and 
four-coordinated sites, respectively. The total Hamiltonian defined upon the underlying lattice 
${\cal L}$ reads:
\begin{eqnarray}
{\mathcal H}_{mix} = - J  \sum_{(i,j) \subset \mathcal J}^{2N} S_{i} \sigma_{j}
     - J' \sum_{(k,l) \subset \mathcal K}^{2N} \sigma_{k} \sigma_{l}
     - D \sum_{i=1}^{N/2} S_{i}^2,     
\label{HD}
\end{eqnarray}
where $\sigma_j = \pm 1/2$ and $S_i = \pm 1/2, \pm 3/2$ are Ising spin variables, $J$ denotes the exchange interaction between nearest-neighbouring spin-1/2 and spin-3/2 pairs and $J'$ labels the interaction between the spin-1/2 pairs that are next-nearest-neighbours on the extended Kagom\'e lattice ${\cal L}$. Finally, the parameter $D$ measures a strength of the uniaxial single-ion anisotropy acting on the spin-3/2 sites and $N$ denotes the total number of the spin-1/2 sites. 

In order to proceed further with calculation, the central spin-3/2 atoms should be firstly decimated from all faces of extended Kagom\'e lattice ${\cal L}$. After the decimation, i.e. after performing a summation over spin degrees of freedom of the spin-3/2 sites (filled circles), the partition function of the mixed-spin 
system can be rewritten as:  
\begin{eqnarray}
{\mathcal Z}_{mix} = \sum_{\{\sigma \}} 
                     \prod_{m=1}^{N/2} \omega_m^{\cal A} (\sigma_i, \sigma_j, \sigma_k, \sigma_l)
                     \prod_{n=1}^{N/2} \omega_n^{\cal B} (\sigma_i, \sigma_j, \sigma_k, \sigma_l). 
\label{ZD}
\end{eqnarray}
Above, the summation is performed over all possible spin configurations available at the spin-1/2 sites and 
the first (second) product is over $N/2$ faces having four spin-1/2 sites $\sigma_i$, $\sigma_j$, $\sigma_k$, $\sigma_l$ placed in the corners of square plaquettes with (without) a central spin-3/2 site in the middle of these plaquettes (see figure \ref{fig1}). The Boltzmann factors $\omega^{\cal A} (a, b, c, d)$ and $\omega^{\cal B} (a, b, c, d)$ assigned to two different kinds of alternating faces, which constitute 
the checkerboard lattice, can be defined as:
\begin{eqnarray}
\omega^{\cal A} (a, b, c, d) \! \! \! &=& \! \! \! 2 \exp[K'(ab + bc + cd + da)/2 + \Delta/4]  
\nonumber \\
\! \! \! && \! \! \!
\bigl \{ \exp(2 \Delta) \cosh[3 K (a + b + c + d)/2] + \cosh[K (a + b + c +d)/2] \bigr \},
\nonumber \\
\omega^{\cal B} (a, b, c, d) \! \! \! &=& \! \! \! \exp[K'(ab + bc + cd + da)/2], 
\label{BF} 
\end{eqnarray}
where $K = J/(k_{\mathrm B} T)$, $K' = J'/(k_{\mathrm B} T)$, $\Delta = D/(k_{\mathrm B} T)$, 
$k_{\mathrm B}$ is Boltzmann's constant, and $T$ stands for the absolute temperature.

\begin{figure}
\vspace{-12mm}
\centerline{\includegraphics[width=1.0\textwidth]{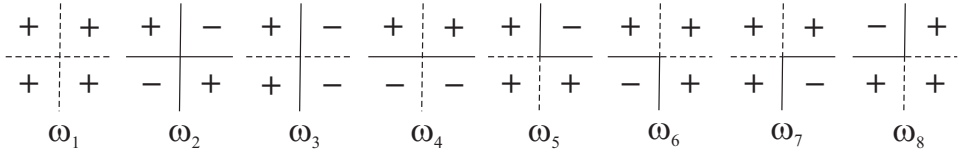}}
\vspace{-13mm}
\caption{The eight possible line arrangements at a vertex of the dual lattice.}
\label{fig2}
\end{figure}

At this stage, the model under investigation can be rather straightforwardly mapped 
onto the staggered eight-vertex model defined on a dual checkerboard lattice ${\cal L_D}$, 
since Boltzmann factors $\omega^{\cal A} (a, b, c, d)$ and $\omega^{\cal B}(a, b, c, d)$ are being 
invariant under the reversal of all four spin variables. Actually, there are maximally eight 
different spin arrangements giving different Boltzmann weights $\omega^{\cal A} (a, b, c, d)$ and 
$\omega^{\cal B}(a, b, c, d)$ for each kind of face.  Diagrammatic representation of eight possible spin arrangements and their corresponding line coverings of the eight-vertex model is shown in figure \ref{fig2}. 
If, and only if, the adjacent spins are aligned opposite to each other, then solid lines are drawn on the edges of the dual 
lattice ${\cal L_D}$, otherwise they are drawn as broken lines.
It can be easily understood that eight possible line coverings around each vertex of the dual 
checkerboard lattice always correspond to two spin configurations, one is being obtained from the other 
by reversing all spins. Since there is even number of solid (broken) lines incident to each vertex of 
the dual lattice ${\cal L_D}$, the model becomes equivalent to the staggered eight-vertex model. 

The Boltzmann weights $\omega^{\cal A} (a, b, c, d)$ and $\omega^{\cal B}(a, b, c, d)$, 
which correspond to eight possible line coverings emerging at vertices of the dual checkerboard lattice, 
can directly be calculated from equation (\ref{BF}): 
\begin{eqnarray}
\omega_1^{\cal A} \! \! \! &=& \! \! \! 2 \exp(K'/2 + \Delta /4)
                               [\exp(2 \Delta) \cosh(3 K) + \cosh(K)], \nonumber \\
\omega_2^{\cal A} \! \! \! &=& \! \! \! 2 \exp(-K'/2 + \Delta/4)[\exp(2 \Delta) + 1], \nonumber \\
\omega_3^{\cal A} \! \! \! &=& \! \! \! \omega_4^{\cal A} = 2 \exp(\Delta/4)[\exp(2 \Delta) + 1], \nonumber \\
\omega_5^{\cal A} \! \! \! &=& \! \! \! \omega_6^{\cal A} = \omega_7^{\cal A} = \omega_8^{\cal A} 
                            = 2 \exp(\Delta/4) [\exp(2 \Delta) \cosh(3 K/2) + \cosh(K/2)]; \label{BW1} \\ 
\omega_1^{\cal B} \! \! \! &=& \! \! \! \exp(K'/2), \qquad \omega_1^{\cal B} = \exp(-K'/2), \nonumber \\
\omega_3^{\cal B} \! \! \! &=& \! \! \! \omega_4^{\cal B} = \omega_5^{\cal B} = \omega_6^{\cal B} = \omega_7^{\cal B} = \omega_8^{\cal B} = 1. 
\label{BW2}
\end{eqnarray}   
Unfortunately, there does not exist general exact solution of the staggered eight-vertex model with arbitrary 
Boltzmann weights $\omega_i^{\cal A}$ and $\omega_j^{\cal B}$ $(i,j=1-8)$. However, if the weights (\ref{BW1}) and (\ref{BW2}) satisfy so-called \textit{free-fermion condition}: 
\begin{eqnarray}
\Omega_1 \Omega_2 + \Omega_3 \Omega_4 = \Omega_5 \Omega_6 + \Omega_7 \Omega_8,
\label{FFC}
\end{eqnarray}  
the staggered eight-vertex model then becomes exactly soluble as the \textit{free-fermion model} solved
several years ago by Hsue, Lin and Wu \cite{Hsu75}. The expressions which enter into the 
free-fermion condition (\ref{FFC}) can be defined through:  
\begin{eqnarray}
\Omega_1 \! \! \! &=& \! \! \! \omega_1^{\cal A} \omega_1^{\cal B} + \omega_2^{\cal A} \omega_2^{\cal B}, \hspace*{3.4cm}
\Omega_2 = \omega_3^{\cal A} \omega_3^{\cal B} + \omega_4^{\cal A} \omega_4^{\cal B}, \nonumber \\
\Omega_3 \! \! \! &=& \! \! \! \omega_5^{\cal A} \omega_6^{\cal B} + \omega_5^{\cal B} \omega_6^{\cal A}, 
\hspace*{3.4cm}
\Omega_4 = \omega_7^{\cal A} \omega_8^{\cal B} + \omega_7^{\cal B} \omega_8^{\cal A}, \nonumber \\
\Omega_5 \Omega_6 \! \! \! &=& \! \! \! 
\omega_1^{\cal A} \omega_1^{\cal B} \omega_3^{\cal A} \omega_3^{\cal B} +
\omega_2^{\cal A} \omega_2^{\cal B} \omega_4^{\cal A} \omega_4^{\cal B} +
\omega_5^{\cal A} \omega_6^{\cal B} \omega_7^{\cal A} \omega_8^{\cal B} +
\omega_5^{\cal B} \omega_6^{\cal A} \omega_7^{\cal B} \omega_8^{\cal A}, \nonumber \\
\Omega_7 \Omega_8 \! \! \! &=& \! \! \! 
\omega_1^{\cal A} \omega_1^{\cal B} \omega_4^{\cal A} \omega_4^{\cal B} +
\omega_2^{\cal A} \omega_2^{\cal B} \omega_3^{\cal A} \omega_3^{\cal B} +
\omega_5^{\cal A} \omega_6^{\cal B} \omega_7^{\cal B} \omega_8^{\cal A} +
\omega_5^{\cal B} \omega_6^{\cal A} \omega_7^{\cal A} \omega_8^{\cal B}.
\label{BWM}
\end{eqnarray}   
It can be readily proved that the free-fermion condition (\ref{FFC}) holds in our case just 
as $D \to \pm \infty$, or $T \to \infty$. The restriction to infinitely strong single-ion anisotropy consequently leads to the familiar phase transitions from the standard Ising universality class, 
because in this case our model effectively reduces to a simple spin-1/2 Ising model on the extended 
Kagom\'e lattice. Within the manifold given by the constraint (\ref{FFC}), the 
free-fermion model becomes critical as long as:
\begin{eqnarray}
\Omega_1 + \Omega_2 + \Omega_3  + \Omega_4 = 2 \mbox{max} \{ \Omega_1, \Omega_2, \Omega_3, \Omega_4 \}.
\label{TCFFC}
\end{eqnarray}
It is noteworthy, however, that the critical condition (\ref{TCFFC}) yields rather reliable 
estimate of the criticality within so-called \textit{free-fermion approximation} \cite{Fan70} 
even if a non-validity of the free-fermion condition (\ref{FFC}) is simply ignored.  

Now, we shall establish an approximate mapping between the staggered and uniform eight-vertex 
models, since the second branch of exact solution is available just for the latter model under 
the \textit{zero-field condition} \cite{Bax82}. For this purpose, let us define 
average Boltzmann weights of the staggered eight-vertex model, which would approximately transform 
the staggered eight-vertex model into the uniform one: 
\begin{eqnarray}
\tilde \omega_i \! \! \! &=& \! \! \! \omega_i^{\cal A} \omega_i^{\cal B}, \qquad (i=1-8). 
\label{BWU}
\end{eqnarray}   
Note that the uniform eight-vertex model satisfies the zero-field condition just 
when its Boltzmann weights are pairwise and symmetrically equal to each other: 
\begin{eqnarray}
\tilde \omega_1 = \tilde \omega_2, \quad \tilde \omega_3 = \tilde \omega_4, \quad 
\tilde \omega_5 = \tilde \omega_6, \quad \tilde \omega_7 = \tilde \omega_8.
\label{S8V}
\end{eqnarray} 
As we already have $\tilde \omega_3 = \tilde \omega_4$, $\tilde \omega_5 = \tilde \omega_6$, 
and $\tilde \omega_7 = \tilde \omega_8$, the zero-field case is consequently reached by 
imposing the condition $\tilde \omega_1 = \tilde \omega_2$ only, or equivalently:
\begin{eqnarray}
\exp(2 \Delta) = \frac{\exp(-2K')-\cosh(K)}{\cosh(3K)-\exp(-2K')},
\label{S8V1}
\end{eqnarray}
According to Baxter's exact solution \cite{Bax82}, the zero-field eight-vertex model 
becomes critical on the manifold (\ref{S8V}) if:
\begin{eqnarray}
\tilde \omega_1 + \tilde \omega_3 + \tilde \omega_5  + \tilde \omega_7 = 
2 \mbox{max} \{\tilde \omega_1, \tilde \omega_3, \tilde \omega_5, \tilde \omega_7 \}.
\label{TCS8V}
\end{eqnarray} 
It is easy to check that $\tilde \omega_1$ represents in our case the largest Boltzmann weight, 
thus, the condition determining the criticality can also be written in this equivalent form:
\begin{eqnarray}
\exp(K'_c) \! \! \! \! \! \! &[& \! \! \! \! \! \! 
\exp(2 \Delta_c) \cosh(3 K_c) + \cosh(K_c)] = \nonumber \\
1 \! \! \! &+& \! \! \! \exp(2 \Delta_c) + 2 \exp(2 \Delta_c) \cosh(3 K_c/2) + 2 \cosh(K_c/2),
\label{TCS8V1}
\end{eqnarray}
where $K_c = J/(k_{\mathrm B} T_c)$, $K'_c = J'/(k_{\mathrm B} T_c)$, $\Delta_c = D/(k_{\mathrm B} T_c)$,
and $T_c$ denotes the critical temperature. It should be stressed, nevertheless, that the critical exponents (with exception of $\delta$ and $\eta$) describing a phase transition of the zero-field eight-vertex model depend on the function 
$\mu = 2 \arctan(\tilde \omega_5 \tilde \omega_7/ \tilde \omega_1 \tilde \omega_3)^{1/2}$, in fact:
\begin{eqnarray}
\alpha = \alpha' = 2 - \frac{\pi}{\mu}, \quad \beta = \frac{\pi}{16 \mu}, \quad \nu = \nu' = \frac{\pi}{2 \mu},
\quad \gamma = \frac{7 \pi}{8 \mu}, \quad \delta = 15, \quad \eta = \frac{1}{4},
\label{CE}
\end{eqnarray} 
Finally, let us explicitly evaluate the critical exponent $\beta$ that determines disappearance 
of the spontaneous order as the critical temperature is approached from below:
\begin{eqnarray}
\beta^{-1} = \frac{32}{\pi} \arctan \biggl\{ \frac{\exp(2 \Delta_c) \cosh(3 K_c/2) + \cosh(K_c/2)}
{[\exp(2 \Delta_c) + 1]^{3/4} [\exp(2 \Delta_c) \cosh(3 K_c) + \cosh(K_c)]^{1/4}} \biggr \}. 
\label{CB}
\end{eqnarray} 
 
\section{Results and discussion}

\begin{figure}
\vspace{-4mm}
\centerline{\includegraphics[width=0.52\textwidth]{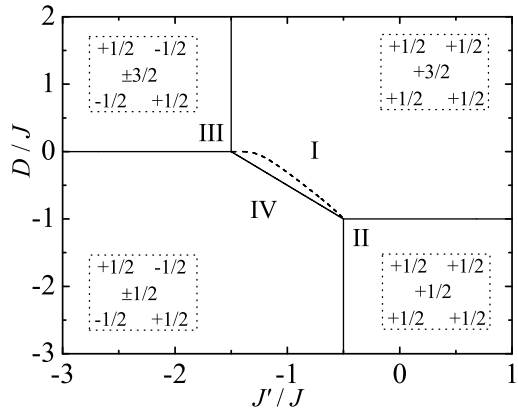}}
\vspace{-6mm}
\caption{Ground-state phase diagram in the $J'-D$ plane when $J > 0$. Broken rectangles schematically 
illustrate a typical spin configuration within each phase. Broken line connecting both triple 
points shows a projection of the critical line (\ref{TCS8V1}) into the $J'-D$ plane.}
\label{fig3}
\end{figure}

Now, let us turn our attention to a discussion of the most interesting results obtained for 
the ground-state and finite-temperature phase diagrams. Solid lines displayed in figure \ref{fig3} 
represent ground-state phase boundaries separating four distinct long-range ordered phases that 
emerge in the ground state when $J > 0$. 
Spin order drawn in broken rectangles shows a typical spin configuration within basic unit cell of each phase. 
As could be expected, a sufficiently strong antiferromagnetic next-nearest-neighbour interaction 
$J'$ alters the structure of the ground state owing to a competing effect with the nearest-neighbour interaction $J$. Due to a competition between the interactions, the central spins are free to flip 
within the phases III and IV and thus, these phases exhibit a remarkable coexistence of order 
and disorder. At last, it is worthwhile to mention 
that a broken line connecting both triple points depicts a projection of the approximate critical line (\ref{TCS8V1}) into the $J'-D$ plane. As this projection crosses zero-temperature plane along the ground-state 
transition line $D/J = -3/2 - J'/J$ between the phases I and IV, it is quite reasonable to suspect 
that this line represent a location of phase transitions between these phases. 

\begin{figure}
\vspace{-2mm}
\centerline{\includegraphics[width=0.49\textwidth]{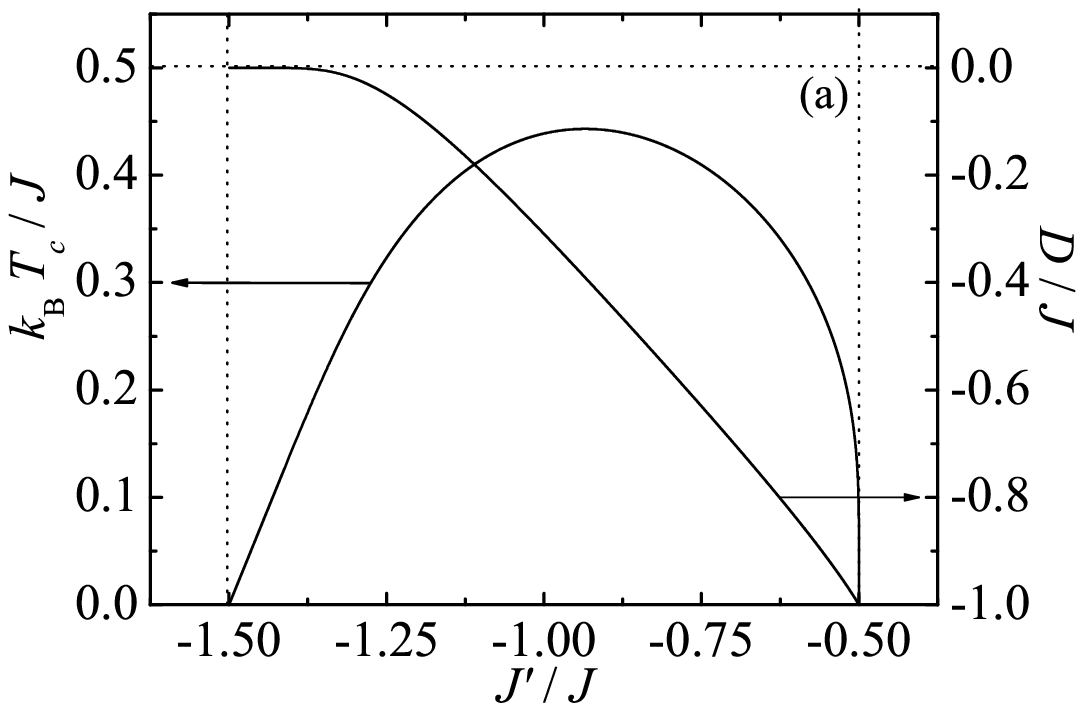} \includegraphics[width=0.49\textwidth]{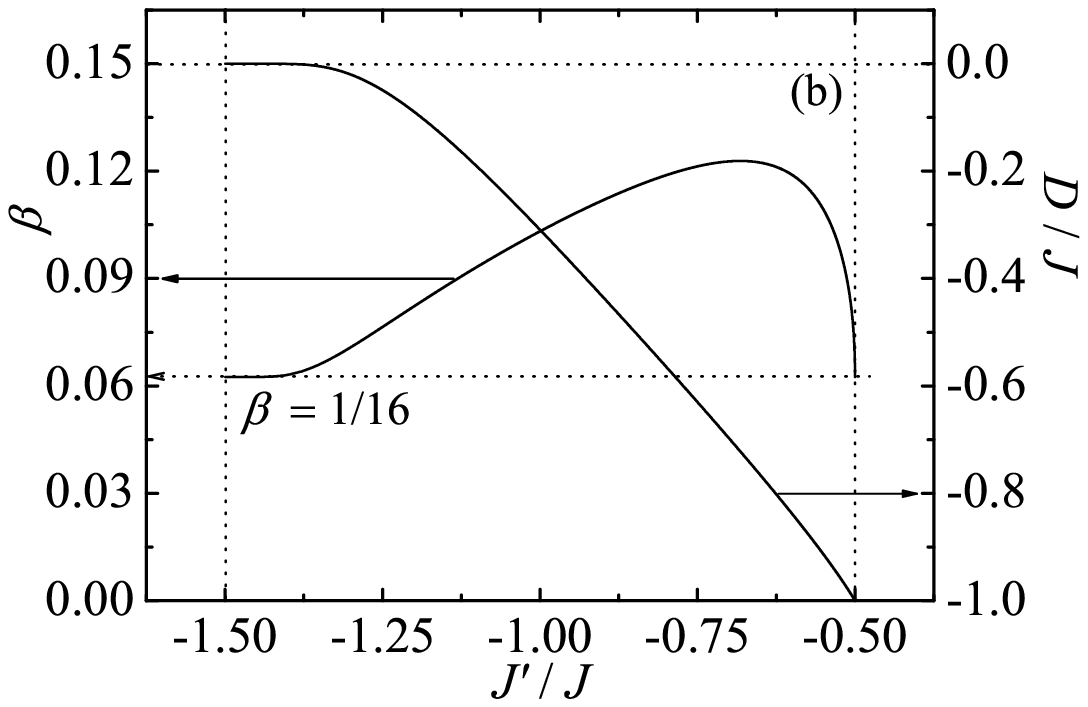}}
\vspace{-5mm}
\caption{(a) The curve scaled to the left axis shows how the critical temperature changes with
a strength of the next-nearest-neighbour coupling $J'/J$, the curve scaled to the right axis depicts 
a variation the single-ion anisotropy parameter along this line; (b) The same as for figure \ref{fig4}(a), 
but the critical exponent $\beta$ is now scaled to the left axis. Broken lines are in both figures 
guides for eyes only.}
\label{fig4}
\end{figure}

Let us investigate more deeply this line of critical points. The critical temperatures calculated 
from the uniform zero-field eight-vertex model must simultaneously obey both the zero-field condition 
(\ref{S8V1}) as well as the critical condition (\ref{TCS8V1}). It is easy to check that the former 
condition necessitates $-1.5 < J'/J < -0.5$ and $-1.0 < D/J < 0.0$. Figure \ref{fig4}(a) displays 
a projection of this critical line into the $J'-T_c$ plane (the dependence scaled to the left axis)
and respectively, a projection into the $J'-D$ plane which is scaled to the right axis. Along this 
critical line, the critical exponents are expected to vary with interaction parameters as they have 
to follow the equations (\ref{CE}). For illustration, figure \ref{fig4}(b) shows how the critical 
index $\beta$ changes along the critical line. Apparently, the exponent $\beta$ approaches its smallest 
possible value $1/16$ by reaching both triple points with zero critical temperature, however, it 
is also quite interesting to ascertain that its greatest value is below the value $1/8$ that 
predicts the universality hypothesis for planar Ising systems \cite{Bax82}. 

\begin{figure}
\vspace{-4mm}
\centerline{\includegraphics[width=0.52\textwidth]{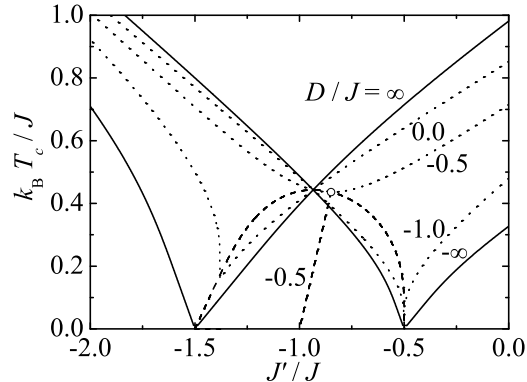}}
\vspace{-6mm}
\caption{A plot critical temperature versus the ratio $J'/J$ for several values of the single-ion 
anisotropy term $D/J$. For details see the text.}
\label{fig5}
\end{figure}

Before concluding, few remarks should be addressed to a global finite-temperature phase diagram plotted 
in figure~\ref{fig5}, which displays the critical temperature as a function of the ratio $J'/J$ for several values of the single-ion anisotropy $D/J$. Critical boundaries depicted as solid lines represent exact critical points obtained from the free-fermion solution (\ref{TCFFC}) of the staggered eight-vertex model
obtained under the constraint (\ref{FFC}), which is fulfilled in the limiting cases $D/J \to \pm \infty$. Dotted critical lines show estimated critical temperatures calculated from the free-fermion approximation simply ignoring a non-validity of the free-fermion condition (\ref{FFC}) for any finite value of $D/J$.
Approximative solution related to the critical points (\ref{TCS8V1}) of the uniform zero-field 
eight-vertex model on the variety (\ref{S8V1}) is displayed as a rounded broken line. 
It is quite obvious from the ground-state phase diagram (figure \ref{fig3}) that a right (left) 
wing of the displayed critical boundaries corresponds to the phase I (III) if $D/J > 0.0$, while it 
corresponds to the phase II (IV) if $D/J < -1.0$. Actually, the exact as well as approximate critical 
points resulting from the free-fermion solution correctly reproduce the ground-state boundaries between 
these phases. When the single-ion anisotropy parameter is selected within the range 
$-1.0 < D/J < 0.0$ (see for instance the curve for $D/J = -0.5$), however, the critical line obtained from the free-fermion approximation meets at a bicritical (circled) point with the critical line of the equivalent 
uniform zero-field eight-vertex model as it has been already reasoned by Lipowski and Horiguchi \cite{Lip95} who have solved similar spin system on the union-jack lattice. In such a case, the right and left part 
(with respect to the bicritical point) of this critical line separate the phases I and IV, respectively, 
and a line of first-order phase transitions is expected to terminate at this special multicritical point. There are strong indications supporting this concept \cite{Lip95}, actually, 
the almost straight broken line depicting the zero-field condition (\ref{S8V1}) should always 
show a coexistence of these two phases as it starts from a point that determines their coexistence 
in the ground state. With regard to the aforementioned arguments one may conclude that a coexistence 
surface between the phases I and IV lies inside the area, which is bounded by the line of bicritical 
points (rounded broken line) having the non-universal interaction-dependent critical exponents. 

Finally, we should remark a feasible appearance of reentrant transitions which can be observed 
in the critical lines nearby the coexistence points $D/J = 0.0$ and $-1.0$. It is quite apparent that 
the observed reentrance can be explained in terms of the coexistence of a partial order and 
partial disorder emerging in both the high-temperature 
reentrant phases III and IV. As a matter of fact, the partial disorder of the spin-3/2 atoms 
can compensate a loss of entropy that occurs in these phases due to a thermally induced 
partial ordering of the spin-1/2 atoms what is in a good accordance with a necessary condition 
conjectured for the appearance of reentrant phase transitions \cite{Aza87}-\cite{Die91}.

\section{Concluding Remarks}

The work reported in the present article provides a relatively precise information on the 
critical behaviour of the mixed spin-1/2 and spin-3/2 Ising model on the extended Kagom\'e lattice 
by establishing a mapping correspondence with the staggered and uniform eight-vertex models, respectively. 
The main focus of the present work has been aimed at the examination of the criticality depending on 
the single-ion anisotropy strength as well as the strength of the competing next-nearest-neighbour interaction. The location of the critical boundaries has accurately been determined from the free-fermion solution of the staggered eight-vertex model and the zero-field solution of the uniform eight-vertex model, respectively, whereas the validity of both mappings is restricted to the certain subspaces of interaction parameters only. In the rest of parameter space, the free-fermion approximation has been used to estimate the critical boundaries as this method should provide meaningful approximation giving rather reliable estimate 
to the true transition temperatures.

The greatest theoretical interest in this model arises due to the remarkable critical line 
consisting of bicritical points, which bounds a coexistence surface between two long-range 
ordered phases. The bicritical points can be characterized by non-universal interaction-dependent 
critical exponents that satisfy the weak universality hypothesis. Moreover, the same arguments 
as those suggested by Lipowski and Horiguchi \cite{Lip95} have enabled us to identify the 
zero-field condition (\ref{S8V1}) with a location of the first-order transition lines separating 
these two ordered phases. 

It should be remarked that the considered spin system also shows reentrant phase transitions on account of 
the competition between the nearest- and next-nearest-neighbour interactions. Our results are in agreement 
with the conjecture \cite{Aza87} stating that the reentrance appears as a consequence of the coexistence 
of a partial order and disorder, namely, the partial disorder induced among spin-3/2 atoms can compensate 
the loss of entropy, which occurs on behalf of the partial ordering of the spin-1/2 atoms in both the high-temperature partially ordered phases.

%
%
\label{last@page}
\end{document}